\documentclass[twocolumn,preprintnumbers,
amsmath,amssymb,aps,prd,nofootinbib,superscriptaddress,
eqsecnum]{revtex4}
\usepackage{graphicx,color}

\makeatletter
\renewcommand{\p@subsection}{}
\makeatother

\begin{document}

\title{Signatures of chiral symmetry restoration in dilepton production}

\author{Chihiro Sasaki}
\affiliation{%
Institute of Theoretical Physics, University of Wroclaw,
PL-50204 Wroclaw,
Poland
}

\date{\today}

\begin{abstract}
We study the structural change of the vector spectral function and integrated production rates of dileptons
in the presence of the chiral mixing induced exclusively at finite density.
The mixing produces multiple bumps and peaks around the vacuum masses of the $\rho, \omega$ and $\phi$ resonances
in the spectral function. The arising modification becomes pronounced when the mass difference between
parity partners decreases.
In particular, the emergent enhancement around the vacuum $\phi$ meson in the production rates serves as an excellent
signature of the partially-restored chiral symmetry in heavy-ion collisions.
\end{abstract}

\maketitle

\section{Introduction}

The role of dynamical chiral symmetry breaking has been extensively explored in the context of
relativistic heavy-ion collisions and the interior of compact stars~\cite{HH,RWvH,HotQCD,FS,nature}.
A number of potential signatures of the restoration of chiral symmetry has been proposed in literature,
whereas no conclusive evidence has been observed in experiment. Dileptons are one of the promising probes
since a virtual photon can propagate in a medium without disturbance. The light vector mesons directly
couple to the electromagnetic current correlator which is the central ingredient in dilepton production.
An enhancement of the dilepton spectra below the $\rho/\omega$ resonances observed at the CERN SPS
is a strong evidence that the vector mesons modify their properties in the medium~\cite{NA60}.

The restoration of chiral symmetry is identified by the current-current correlation functions degenerate
in opposite parity channels. It is characteristic in hot/dense matter that the pionic interaction
yields a mixing of the vector with the axial-vector correlator. At low temperature or density, this
is expressed as a model-independent theorem~\cite{DEI,Krippa98,Chanfray}. Some systematic calculations at finite temperature
exist via the theorem~\cite{MHW,UBW} and in a chiral reduction formulism~\cite{DTZ}.
In Ref.~\cite{HSWhot}, it was shown
that at finite temperature the chiral symmetry restoration forces the chiral mixing to vanish within
a chiral effective theory of the $\rho$, $a_1$ mesons and the pion. The same trend is found in Ref.~\cite{HRhot}
where the in-medium axial-vector spectral function was constructed, via the Weinberg sum rules~\cite{WSR}, from the vector
current correlator that describes the dilepton data. There a substantial mass drop of the $a_1$ to the
$\rho$ mass and a width broadening toward the chiral symmetry restoration were reported.

In contrast to the above mixing induced by pion loops, there exists a new class of the chiral mixing that modifies
the dispersion relations of the vector and axial-vector mesons at finite density.
This mixing was shown to emerge in the low-energy effective theory of QCD based on the AdS/CFT correspondence~\cite{DHholo}.
Its phenomenological impact in dense QCD matter is a significant change of the vector spectrum which may
lead to multiple bumps and peaks in dilepton rates at a few times of the normal nuclear matter density
whereas assumed that the mesons do no change their masses~\cite{HSdense}.

In this paper, we introduce the order parameter as a function of temperature and chemical potential to
the vector current correlator in the presence of the density-induced mixing and study a possible signal
of partial restoration of chiral symmetry to be verified in dilepton measurement. A special emphasis is
put on the competition between the chiral mixing and the mass degeneracy of the parity partners.
It is shown that decreasing the mass difference reinforces the structural change of the spectral function
and this persists in the integrated dileption rate even with a marginal strength of the mixing.

\section{Chiral mixing in dense matter}

At finite baryon density, charge-conjugation invariance is explicitly violated
whereas parity invariance remains intact. The chiral Lagrangian thus includes a term
\begin{equation}
{\mathcal L}_{\rm mix} = 2c\,\epsilon^{0\mu\nu\lambda}\mbox{tr}\left[
\partial_\mu V_\nu\cdot A_\lambda + \partial_\mu A_\nu\cdot V_\lambda
\right]\,,
\label{lag1}
\end{equation}
for the vector $V_\mu$ and the axial-vector $A_\mu$ mesons with the total anti-symmetric tensor
$\epsilon^{0123}=1$ as well as a mixing parameter $c$.
The mixing yields the modified dispersion relations for the transverse polarizations~\cite{DHholo}
\begin{equation}
p_0^2 - \vec{p}^2
= \frac{1}{2}\left[
m_V^2 + m_A^2 \pm \sqrt{(m_A^2 - m_V^2)^2 + 16 c^2 \vec{p}^2}
\right]\,,
\label{disp}
\end{equation}
with the lower sign for the vector and the upper one for the axial-vector mesons.
The longitudinal modes obey the standard dispersion relation, $p_0^2 - \vec{p}^2 = m_{V,A}^2$.

In a model based on AdS/CFT at finite baryon chemical potential~\cite{DHholo}, the mixing strength
$c$ possesses an explicit dependence on the baryon density $\rho_B$ and takes a rather large value
$c \simeq 1$ GeV at the normal nuclear matter density $\rho_0$. This results in the onset of
vector condensation at a density slightly above $\rho_0$, which is an apparent drawback of large $N_c$
since the known property of nuclear matter excludes this possibility under a realistic setup with $N_c=3$.
This strongly suggests a non-trivial contribution as $1/N_c$ corrections, which may change the mixing
strength quantitatively. The value of $c$ can be determined in the standard chiral approach by
replacing Eq.~(\ref{lag1}) with the $\omega\rho a_1$ term that arises from the gauged Wess-Zumino-Witten (WZW) action~\cite{KM}
\begin{equation}
{\mathcal L}_{\omega\rho a_1}
= g_{\omega\rho a_1}\langle\omega_0\rangle\epsilon^{0\mu\nu\lambda}
\mbox{tr}\left[
\partial_\mu V_\nu\cdot A_\lambda + \partial_\mu A_\nu\cdot V_\lambda
\right]\,,
\end{equation}
where the iso-scalar $\omega$ field is replaced with its expectation value,
$\langle\omega_0\rangle = g_{\omega NN}\cdot\rho_B/m_\omega^2$, obtained in the conventional Walecka model.
With empirical numbers, one finds the mixing strength $c = g_{\omega\rho a_1}\langle\omega_0\rangle \simeq 0.1$ GeV
at $\rho_0$~\cite{HSdense}. Although the weak mixing has little importance in the vector-current correlation function at $\rho_0$,
a distinct modification emerges at higher density leading to a stronger mixing $c \simeq 0.3$ GeV in the correlator and
consequently in production rates of a lepton pair~\cite{HSdense}.
Note that no further term that changes the dispersion relation appears from the WZW action by replacing $\omega_\mu$
with its expectation value. Induced interactions will modify the masses and widths via loop diagrams.

Whereas the mixing effect is expected to become more important at higher density, a crucial question to be answered is
how the signals of the partial restoration of chiral symmetry would emerge in observables. Expanding the dispersion relation
(\ref{disp}) for a small momentum $\vec{p}$, we readily see
\begin{equation}
p_0^2 \simeq m_{A,V}^2 + \left(1 \pm \frac{4c^2}{m_A^2 - m_V^2}\right)\vec{p}^2\,,
\end{equation}
where the second term in the parentheses becomes enhanced as the mass difference between the parity partners decreases
even with a marginal strength $c$.
The above expression breaks down when the mass splitting $m_A^2-m_V^2$ becomes small, so that a self-consistent determination of
both $m_{A,V}$ and $c$ is necessary to get a qualitative insight into the relevance of the mixing effect.

The mixing strength $c$ and the meson masses $m_{V,A}$ in general carry certain medium effects
that are inherently related to each other, and it requires a suitable extension of the previous study
to determine those parameters in a self-consistent way.
We begin with the vector and axial-vector current correlators in matter
\begin{equation}
G_{V,A}^{\mu\nu}(p_0,\vec{p})
= P_L^{\mu\nu}G_{V,A}^L(p_0,\vec{p}) + P_T^{\mu\nu}G_{V,A}^T(p_0,\vec{p})\,,
\end{equation}
with the polarization tensors
\begin{eqnarray}
P_{T,\mu\nu}
&=&
g_{\mu i}\left(\delta_{ij}-\frac{\vec{p}_i\vec{p}_j}{\vec{p}^2}\right)g_{j\nu}\,,
\nonumber\\
P_{L,\mu\nu}
&=&
-\left(g_{\mu\nu}-\frac{p_\mu p_\nu}{p^2}\right) - P_{T,\mu\nu}\,.
\end{eqnarray}
The longitudinal and transverse parts are expressed as~\cite{HSdense}
\begin{eqnarray}
&&
G_V^L = \left(\frac{g_V}{m_V}\right)^2\frac{-s}{D_V}\,,
\quad
G_V^T = \left(\frac{g_V}{m_V}\right)^2\frac{-sD_A + 4c^2\vec{p}^2}{D_VD_A - 4c^2\vec{p}^2}\,,
\nonumber\\
&&
G_A^L = \left(\frac{g_A}{m_A}\right)^2\frac{-s}{D_A}\,,
\quad
G_A^T = \left(\frac{g_A}{m_A}\right)^2\frac{-sD_V + 4c^2\vec{p}^2}{D_VD_A - 4c^2\vec{p}^2}\,,
\nonumber\\
\label{spectra}
\end{eqnarray}
with $s = p_0^2 - \vec{p}^2$ and the coupling of the vector/axial-vector meson to the
corresponding current $g_{V,A}$ as well as the propagator inverse without the mixing
$D_{V,A} = s - m_{V,A}^2 + im_{V,A}\Gamma_{V,A}(s)$.
The spin-averaged correlators are defined by $G_{V,A} = \frac{1}{3}\left(G_{V,A}^L + 2G_{V,A}^T\right)$.
The labels $(V,A)$ refer to the iso-vector states $(\rho,a_1)$ and
the iso-singlet states $(\omega,f_1(1285))$ and $(\phi,f_1(1420))$.

Phenomenology of the pseudo-scalar and vector mesons is well described in the non-linear chiral
Lagrangian based on the generalized hidden local symmetry (GHLS)~\cite{BKY,KM,HSghls}.
The vector and axial-vector meson masses are related via the pion decay constant $f_\pi$ as
\begin{equation}
m_A^2 - m_V^2 = g^2\frac{m_A^2}{m_V^2}f_\pi^2 \equiv \delta m^2\,,
\end{equation}
with the gauge coupling $g$. Thus, the mass difference $\delta m$ serves as the order parameter of
spontaneous chiral symmetry breaking.
In order to introduce a non-trivial medium effect that induces
a chiral phase transition, we shall replace $f_\pi$ with the in-medium expectation value of the sigma
field $\langle\sigma\rangle$ computed in the standard linear sigma model.
We note that vanishing $\delta m$ leads to $m_A \to m_V$, $\Gamma_A \to \Gamma_V$ and $g_A \to g_V$,
so that the spectral functions become identical in the vector and axial-vector channels for non-vanishing $c$.

\begin{figure}
\includegraphics[width=8cm]{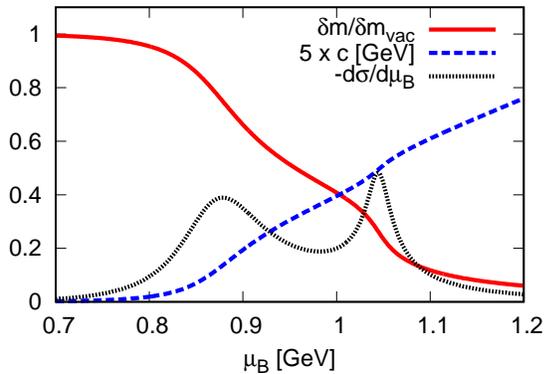}
\caption{The mass difference between the vector and axial-vector mesons $\delta m$ normalized its vacuum value,
the chiral mixing $c$ and the $\mu_B$-derivative of the order parameter
at fixed temperature $T=50$ MeV as functions of baryon chemical potential $\mu_B$.
The $\mu_B$ profiles of
the two mean fields, $\langle\sigma_q\rangle$ and $\langle\omega_0\rangle$ are taken from \cite{Marczenko}}.
\label{para}
\end{figure}

For an illustrative calculation, we employ the nucleon parity-doublet model to obtain
medium profiles of the sigma and omega expectation values in the mean field approximation, where
the lowest nucleon and its negative-parity counterpart $N(1535)$ play the essential role in describing the nuclear
ground state~\cite{pdm,SMpdm}. The model has been extended further by introducing the confinement nature on top of
the chiral dynamics and confronted with the properties of neutron stars~\cite{Marczenko}.
In Fig.~\ref{para}, we show the resultant mass-splitting and the chiral mixing at temperature $T=50$ MeV
as functions of baryon chemical potential $\mu_B$.
It is characteristic of the nucleon parity doublet model that the VEV of the sigma field drops twice, at the liquid-gas (LG) transition
and at the chiral phase transition/crossover at zero temperature~\cite{pdm}. When the model is applied to finite temperature, the first-order
LG transition becomes a crossover via a second-order critical point. One still sees its remnant in the VEV as an abrupt change at a relatively
low chemical potential $\mu_B$~\cite{SMpdm}.
The model with the parameters constrained by the maximum masses and the compactness of the neutron stars in~\cite{Marczenko}
yields a chiral crossover at $\mu_B \simeq 1.05$ GeV (equivalently the net baryon density $\rho_B \simeq 2.5\,\rho_0$),
at which the lowest nucleon and the $N(1535)$ become nearly degenerate.
That way, we can handle the low-lying parity doublers
both in the meson and baryon sectors on the equal footing.

In the subsequent calculations, we shall set the vector meson mass to its vacuum value at any $T$ and $\mu_B$
for simplicity, and this prescription is supported in line with the known phenomenology as well~\cite{RWvH}.
Thus, the axial-vector mass $m_A = \sqrt{m_V^2 + \delta m^2}$ and the decay width $\Gamma_A(m_V,m_A)$
vary with respect to $T$ and $\mu_B$ according to $\delta m$ in the spectral function (\ref{spectra}).
Also, we use the form of the vacuum decay widths for the $\rho$ and $a_1$ mesons~\cite{HSdense}
\begin{eqnarray}
\Gamma_\rho(s)
&=&
\Theta(s-4m_\pi^2)\frac{m_\rho}{\sqrt{s}}\left(\frac{s-4m_\pi^2}{m_\rho^2-4m_\pi^2}\right)^{3/2}\Gamma_\rho\,,
\nonumber\\
\Gamma_\rho
&=&
\frac{1}{6\pi m_\rho^2}\left(\frac{m_\rho^2-4m_\pi^2}{4}\right)^{3/2}g_{\rho\pi\pi}^2\,,
\nonumber\\
\Gamma_{a_1}(s)
&=&
\Theta(s-(m_\rho+m_\pi)^2)\frac{g^2}{8\pi m_{a_1}^2}\left[
g^2 f_\pi^2 + \frac{m_{a_1}^2-m_\rho^2}{12 m_{a_1}^2}s
\right.
\nonumber\\
&&\times
\left.
\left(1-\frac{(m_\rho+m_\pi)^2}{s}\right)
\left(1-\frac{(m_\rho-m_\pi)^2}{s}\right)
\right]
\nonumber\\
&&\times
\sqrt{1-\frac{(m_\rho+m_\pi)^2}{2}}
\sqrt{1-\frac{(m_\rho-m_\pi)^2}{2}}\,,
\label{width}
\end{eqnarray}
with the parameter $g=6.61$~\cite{HSWhot}.
We take the following values at $T=\mu_B=0$ for our calculations:  $f_\pi=92.4$ MeV, $m_\pi=0.14$ GeV,
$m_\rho=0.77$ GeV, $g_\rho = 0.119$ GeV$^2$, $g_{\rho\pi\pi} = 6$ and $m_{a_1}=1.26$ GeV~\cite{PDG},
which lead to the on-shell decay widths, $\Gamma_\rho(s=m_\rho^2)=0.15$ GeV and $\Gamma_{a_1}(s=m_{a_1}^2)=0.33$ GeV.

In Eq.(\ref{width}) we deal with the mass $m_{a_1}$ and the order parameter $f_\pi=\langle\sigma\rangle$ as in-medium quantities,
thus the width $\Gamma_{a_1}$ carries the medium effect corresponding to partial restoration of the chiral symmetry. One readily finds
that the $\Gamma_{a_1}$ vanishes as $f_\pi \to 0$, consistently to the $a_1$ carrying the equal mass to the $\rho$ meson when
the chiral symmetry gets restored. Therefore, in order to ensure the degenerate current correlators, $G_V=G_A$, we must include
an additional term to Eq.~(\ref{width}) which contributes to the total $a_1$ width with the equal strength to the $\Gamma_\rho$.
Here the scalar degree of freedom comes in. In the linear sigma model, the axial-vector meson decays into the lowest scalar meson
$\sigma$ and pion. At the chiral restoration, the scalar meson mass becomes equal to the pion mass, so that the decay rate is
expected to be $\Gamma_{a_1 \to \sigma\pi} = \Gamma_{\rho \to 2\pi}$. Therefore, we adopt the schematic parameterizations
\begin{eqnarray}
\Gamma_{a_1}(s)
&=&
\Gamma_{a_1 \to \rho\pi}(s) + \delta\Gamma_{a_1}(s)\,,
\nonumber\\
\delta\Gamma_{a_1}(s)
&=&
\Theta(s-4m_\pi^2)\Gamma_\rho\left[
1 - \left(\frac{\langle\sigma\rangle}{\langle\sigma\rangle_{\rm vac}}\right)^2
\right]\,.
\label{dgam}
\end{eqnarray}

The chiral mixing between $\omega$-$f_1(1285)$ and $\phi$-$f_1(1420)$ can be introduced in a straightforward way~\cite{HSdense}.
We employ the constant-width approximation of narrow peaked mesons above threshold:
$\Gamma_\omega=8.49$ MeV, $\Gamma_\phi=4.26$ MeV, $\Gamma_{f_1(1285)}=24.3$ MeV and
$\Gamma_{f_1(1420)}=54.9$ MeV~\cite{PDG}, with the kinematical constraints
\begin{eqnarray}
\Gamma_\omega(s)
&=&
\Theta(s-9m_\pi^2)\Gamma_\omega\,,
\nonumber\\
\Gamma_\phi(s)
&=&
\Theta(s-4m_K^2)\Gamma_\phi\,,
\nonumber\\
\Gamma_{f_1(1285)}(s)
&=&
\Theta(s-16m_\pi^2)\Gamma_{f_1(1285)}\,,
\nonumber\\
\Gamma_{f_1(1420)}(s)
&=&
\Theta(s-(m_\pi+2m_K)^2)\Gamma_{f_1(1420)}\,.
\nonumber\\
\end{eqnarray}
The coupling constants to the vector current are related to $g_\rho$ via chiral symmetry, so that
\begin{equation}
g_\omega = \frac{1}{3}\frac{m_\omega^2}{m_\rho^2}g_\rho\,,
\quad
g_\phi = \frac{\sqrt{2}}{3}\frac{m_\phi^2}{m_\rho^2}g_\rho\,.
\end{equation}
We employ the same parameterizations for $\delta\Gamma_{f_1}$ as in Eq.~(\ref{dgam}).

It may not be appropriate to naively replace $\langle\omega_0\rangle$ with $\langle\phi_0\rangle$
to obtain the mixing between the $\phi$ and $f_1(1420)$ states since the mean-field calculations typically
yield a $\langle\bar{s}s\rangle$ decreasing in a much milder way than the light-quark condensate.
This results in a significant delay of vanishing $\delta m$ which contradicts the vector screening mass
of the $\bar{s}s$ state seen in lattice simulations~\cite{latscr}. There the screening masses exhibit a substantial modification
at around the pseudo-critical temperature and this arises nearly independent of the quark-flavor content. Therefore, we will not proceed to
a three-flavored parity doublet model, but rather impose the same critical behavior of the meson masses as the lattice
observation. To this end, the simplest way is to assume that the modification of $\delta m$ for the $\phi$ and $f_1(1420)$
is dominated by the two-flavor physics. In the subsequent calculations, we will set at the chiral crossover $\delta m/\delta m_{\rm vac}=0.36$
for the strange vector mesons, whereas the two-flavored parity doublet model yields $\delta m/\delta m_{\rm vac}=0.26$ for the light vector states.
The same re-scaling is applied to the expectation value of $\langle\phi_0\rangle$.

\section{Vector spectral function}

To illustrate the competition between the density-induced mixing and chiral symmetry restoration,
we define the vector spectral function as the imaginary part of the current correlator, $\mbox{Im}G_V$,
for a given three-momentum $\vec{p}$. The characteristic feature with the chiral mixing is that the longitudinal
part of the correlator is peaked at the vacuum mass of the vector state, whereas the transverse parts are
modified by the transverse vector-meson with a shifted mass downward and the transverse axial-vector with
a shifted mass upward. Consequently, the vector spectral function exhibits three bumps in hadronic phase
if the mixing strength is sufficiently large~\cite{HSdense}.
As the system gradually approaches the chiral symmetry restoration in dense matter, those bumps and peaks
are supposed to change their locations according to the dispersion relations~Eq.(\ref{disp}).

\begin{figure*}
\includegraphics[width=8cm]{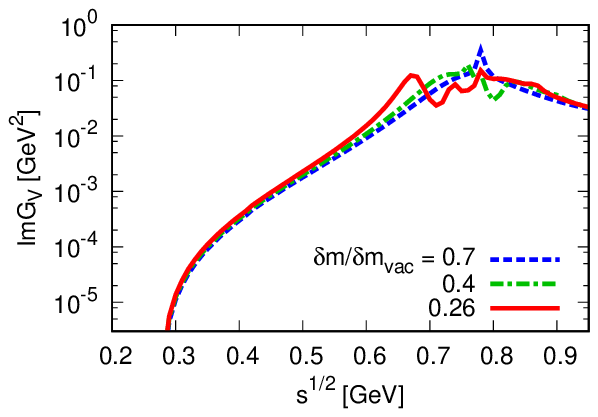}
\includegraphics[width=8cm]{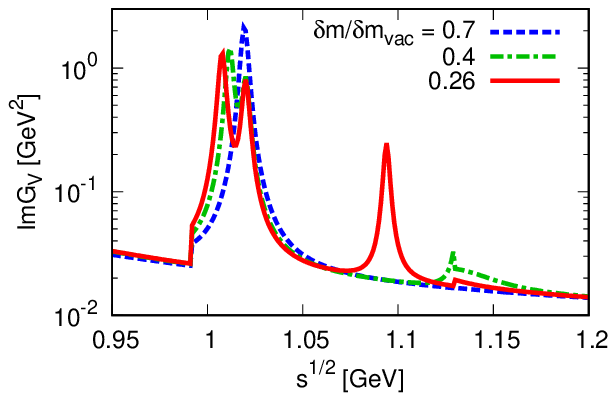}
\caption{The vector spectral function with $|\vec{p}|=0.5$ GeV at $T=50$ MeV in the $\rho$-$\omega$ (left)
and $\phi$ (right) channels for $\delta m/\delta m_{\rm vac} = 0.7, 0.4$ and $0.26$.}
\label{SP}
\end{figure*}

In Fig.~\ref{SP} we show the vector spectral function at $T=50$ MeV in the $\rho$-$\omega$ and $\phi$ channels
using Eq.~(\ref{spectra}) with $|\vec{p}|=0.5$ GeV and the condensates shown in Fig.~\ref{para}.
In the $\rho$-$\omega$ channel, the distinct peak of the longitudinal $\omega$ meson stays at any $\mu_B$ whereas
its strength is decreased because of the mixing effect. At $\delta m/\delta m_{\rm vac} =0.7$, the system remains far from the chiral
symmetry restoration and the mixing effect with $c \simeq 34$ MeV is totally irrelevant. At higher $\mu_B$, the mixing
sets in because of decreasing $\delta m/\delta m_{\rm vac}$ and the spectrum exhibits multiple bumps of the transverse polarizations, which changes the locations in $\mu_B$.
Especially, the transverse $\rho$ yields a substantial contribution near the chiral crossover and the spectrum becomes
enhanced at small $\sqrt{s}$.
The aforementioned structure with the three bumps is best preserved in the $\phi$ channel at $\delta m/\delta m_{\rm vac}=0.26$.
The most-left peak of the transverse $\phi$ shifts its position to lower $\sqrt{s}$ as $\mu_B$ is increased.
At some point, it meets the threshold and becomes cut off from the spectrum. This will lead to a two-peak structure
at a smaller $\delta m/\delta m_{\rm vac}$.
We emphasize that the distinct modification in the spectral function disappears when the masses of the axial-vector mesons
are frozen to be constant, since the mixing $c$ is not very strong at any $\mu_B$ considered in this study and is insufficient to modify
the propagators. Therefore, the drastic change seen in Fig.~\ref{SP} is the direct consequence of the chiral symmetry restoration.

\begin{figure*}
\includegraphics[width=8cm]{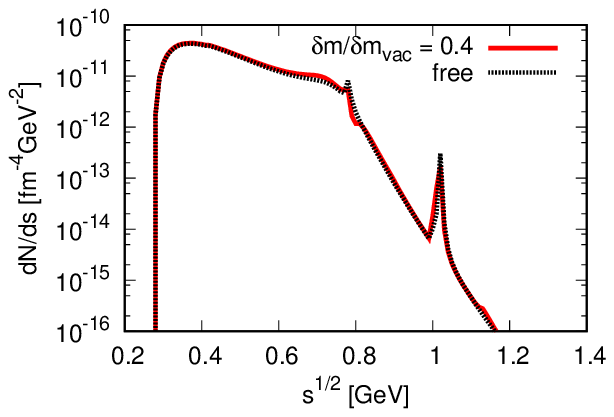}
\includegraphics[width=8cm]{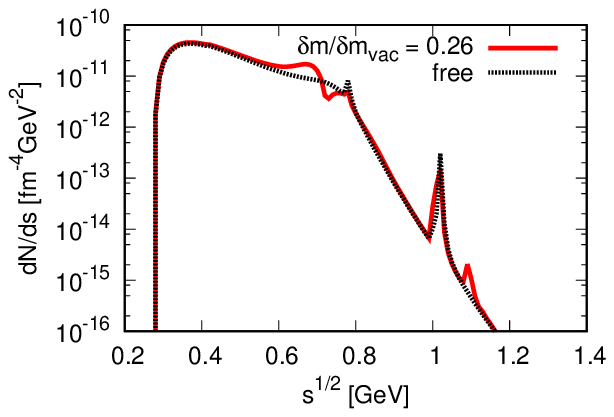}
\caption{The dilepton production rate at $T=50$ MeV for $\delta m/\delta m_{\rm vac}=0.4$ (left) and $0.26$ (right).}
\label{DL}
\end{figure*}

It is a straightforward and intriguing application to calculate the production rate of a lepton pair emitted from
dense matter via a virtual photon. The differential rate in a medium at finite $T$ and $\mu_B$ is given
in terms of the imaginary part of the vector current correlator~\cite{RWvH}
\begin{equation}
\frac{dN}{d^4p}(p_0,\vec{p};T,\mu_B)
= \frac{\alpha^2}{\pi^3 s}\frac{\mbox{Im}G_V(p_0,\vec{p};T,\mu_B)}{e^{p_0/T}-1}\,,
\end{equation}
with $\alpha=e^2/4\pi$ the electromagnetic coupling constant.
The three-moemntum integrated rate is given by
\begin{equation}
\frac{dN}{ds}(s;T,\mu_B)
= \int\frac{d^3\vec{p}}{2p_0}\frac{dN}{d^4p}(p_0,\vec{p};T,\mu_B)\,.
\end{equation}
In Fig.~\ref{DL}, the integrated rate in the range of $0 \leq |\vec{p}| \leq 1$ GeV at $T=50$ MeV is presented
for $\mu_B/\mu_B^c = 0.75$ and $1.0$.
The characteristic structure remains there even after the $\vec{p}$-integral is performed, although somewhat weakened.
The modified $\rho$ and $\omega$ mesons yield a significant enhancement in low $\sqrt{s}$ region, and
the axial-vector counterpart of the $\phi$ meson produces one additional peak around $\sqrt{s}\sim 1.1$ GeV near the chiral symmetry restoration.
The transverse $\phi$ state appears rather close to the longitudinal polarization, and thus the entire spectrum
becomes broadened.

In the above calculations, in-medium broadening of the widths were not taken into account.
The presence of hot and dense matter strongly modifies the shape of the vector spectral function~\cite{RWvH}.
In fact, a systematic treatment of the in-medium $\rho$ meson successfully describes the dimuon data
of the NA60 Collaboration in heavy-ion collisions at CERN SPS~\cite{vHR}, where the characteristic baryon-induced
interactions play the central role. Thus, in a more realistic calculation with the chiral mixing, such broadened
widths may screen the additional bumpy structure of the spectrum, especially in the $\rho$-$\omega$ channel.
On the other hand, the $\phi$-meson spectrum receives much weaker modifications and the $\phi$
remains a well-defined narrow resonance in a medium~\cite{Rapp}. We therefore anticipate that the modification
of the vector spectrum in the $\phi$ channel serves as a convincing signal of the chiral symmetry restoration.
The additional peaks of the modified $\phi$ and $f_1(1420)$ states will lead to the two bumps,
one is below and another is above the vacuum $\phi$ meson mass,
in the dilepton rates by summing up the evolution history of the created matter.
We note that the transverse polarizations change their masses systematically to lower values
as $\mu_B$ increases, and at some point the lower peak is cut off at the $2m_K$ threshold.
Therefore, the number of bumps actually seen in the rate depends on the chemical potential $\mu_B$.

\begin{figure}
\includegraphics[width=8cm]{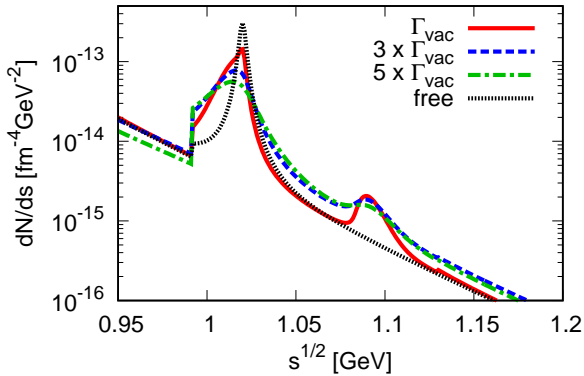}
\caption{The dilepton production rate at $T=50$ MeV for $\delta m/\delta m_{\rm vac}=0.26$ with various decay widths of the $\phi$ meson.}
\label{broadening}
\end{figure}

As an illustration, we show the rate at the chiral restoration point $\delta m/\delta m_{\rm vac}=0.26$ with a larger width of the $\phi$ meson
than its vacuum value by factors of $3$ and $5$ in Fig.~\ref{broadening}.
The peak of the longitudinal polarization is clearly suppressed and the entire shape is much broadened, as expected.
Whereas, the enhancement below and above the vacuum $\phi$ meson clearly survives. Thus, there remains a good chance
to observe a trace of the chiral symmetry restoration via the chiral mixing in more realistic calculations of the dilepton rate 
if the matter is sufficiently dense and cold.

\section{Conclusions}

We have studied the consequences of the density-induced chiral mixing in the vector spectral function
and the dilepton production rate, with a special emphasis on ensuring the restoration of chiral symmetry.
The absence of charge-conjugation invariance at finite chemical potential naturally leads to
the modified dispersion relations for the transverse polarizations of the vector and axial-vector mesons.
Multiple bumps and peaks arise in the spectral function and they gradually change the locations according to
the onset of the chiral symmetry restoration.

It is striking that, even with a marginal strength of the mixing, $c \sim 0.1$ GeV, decreasing the order parameter $\delta m$
reinforces the structural change of the spectral function. This is in a sharp contract to the scenario without
the mass degeneracy~\cite{HSdense}. Therefore, the arising enhancement around the $\rho,\omega$ and $\phi$ resonances in the dilepton rates
serves as a clear signature of the restored chiral symmetry. Especially, the $\phi$ meson and its axial-vector counterpart
carry much more promising signals than the $\rho$ and $a_1$ states since the $\phi$ likely remains a well-defined
narrow resonance in a hot/dense medium. This will be an excellent signal of the chiral symmetry restoration
to be verified in heavy-ion collisions at FAIR, NICA and J-PARC.

In our calculations, the in-medium widths of the axial-vector mesons were minimally modified such that the current
correlation functions in the vector and axial-vector channels coincide at the restoration point.
Hadronic many-body approaches will certainly modify the spectral function quantitatively, and it is mandatory
to determine the axial-vector decay rates more precisely near the chiral symmetry restoration on top of the chiral mixing.
Despite the simplifications made in our calculations, the strategy does not rely on the detailed prescriptions to handle
the many-body dynamics. In fact, the two current correlators must coincide at the restoration, $G_V=G_A$,
with the total in-medium widths $\Gamma_{V,A}^\ast$ independently on how they are evaluated.
Further study on the vector spectral function including major baryon-induced effects and the production of the $\phi$ meson
in dense matter is a work in progress.

\subsection*{Acknowledgments}

I acknowledge stimulating discussions with
W.~Broniowski, T.~Galatyuk, K.~Redlich, P.~Salabura and N.~Xu.
I also thank Michal Marczenko for providing me the condensates calculated in his model. 
This work has been partly supported
by the Polish Science Foundation (NCN) under
Maestro Grant No. DEC-2013/10/A/ST2/00106
and OPUS Grant No. 2018/31/B/ST2/01663.


\end{document}